# Ultrafast 4D Scanning Transmission Electron Microscopy for Imaging of Localized Optical Fields

Petr Koutenský,* Neli Laštovičková Streshkova, Kamila Moriová, Marius Constantin Chirita Mihaila, Alexandr Knápek, Daniel Burda, and Martin Kozák*

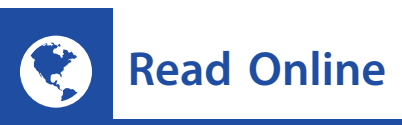

Read Online

ACCESS | Metrics & More | Article Recommendations | Supporting Information

**ABSTRACT:** Ultrafast electron microscopy aims for imaging transient phenomena occurring on nanoscale. One of its goals is to visualize localized optical and plasmonic modes generated by coherent excitation in the vicinity of various types of nanostructures. Such imaging capability was enabled by photon-induced near-field optical microscopy, which is based on spectral filtering of electrons inelastically scattered due to the stimulated interaction with the near-field. Here, we report on the development of ultrafast four-dimensional (4D) scanning transmission electron microscopy, which allows us to image the transverse components of the optical near-field while avoiding the need of electron spectral filtering. We demonstrate that this method is capable of imaging the integrated Lorentz force generated by optical near-fields of a tungsten nanotip and the ponderomotive potential of an optical standing wave with a spatial resolution of 21 nm.

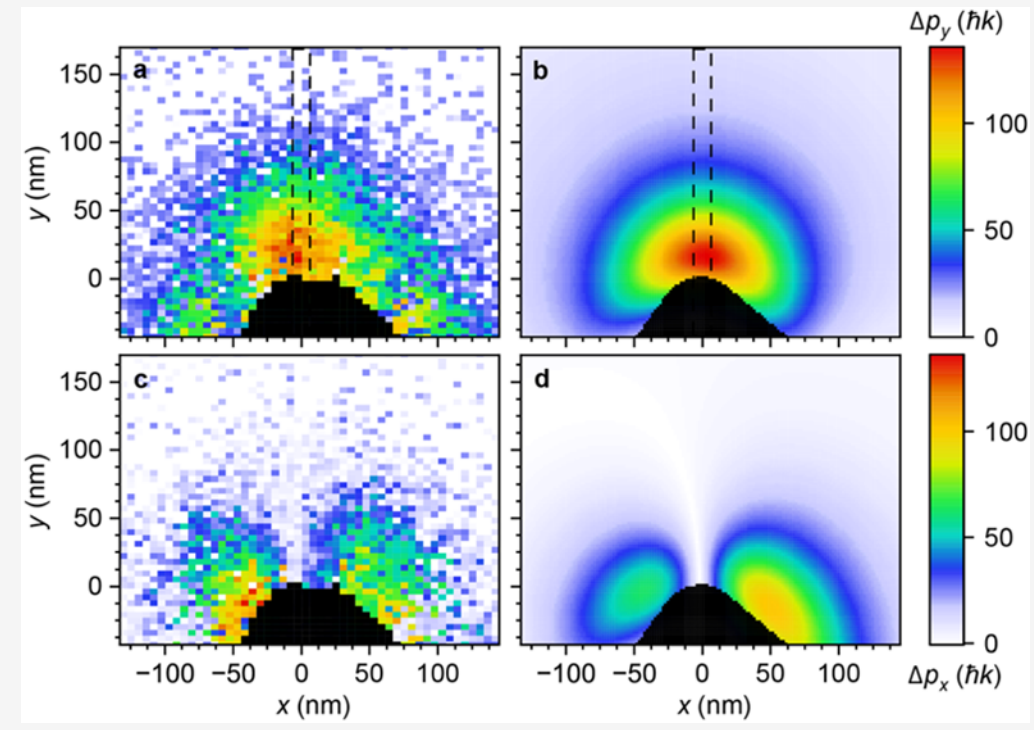

## INTRODUCTION

Electron microscopy has evolved into a versatile tool that provides insight into the nanoworld. Image formation in electron microscopes is mostly based on elastic scattering of electrons on the electrostatic potential of atomic cores while electron spectroscopy relies on inelastic scattering of electrons interacting with shell electrons in the specimen atoms or spontaneously exciting plasmons, phonons or other types of excitations in the studied material. Recent developments have enabled, e.g., to perform vibrational spectroscopy of individual atoms,[1,2] to visualize complex structure of viruses or proteins[3,4] or to measure spatially resolved electron scattering patterns using a general method of four-dimensional (4D) scanning transmission electron microscopy.[5] In addition to the ultimate spatial and spectral resolution provided by the state-of-the-art electron microscopes, an insight into ultrafast dynamics occurring on femtosecond to nanosecond time scales has been enabled by combining pulsed laser and electron beams in a pump−probe fashion.[6−9]

One of the fundamental dynamical processes occurring on subnanometer spatial scales and femtosecond time scales is light-matter interaction. When a material is illuminated by a light wave, the oscillatory motion of electrons with respect to heavy static ions leads to formation of oscillating dipoles which emit the electromagnetic field at the frequency of the incident field. In homogeneous materials, this effect gives rise to a refractive index because of the phase delay of the emitted wave, whose interference with the incident wave slows down the phase velocity of light. When the oscillating dipoles are excited in a nanostructure with spatial dimensions smaller than the wavelength of the excitation light field, the superposition of the dipole-like radiation emitted by the nanostructure with the incident wave leads to a local enhancement of the electric field amplitude of the field oscillating at the optical frequency. Such local field enhancement has many applications, for example, surpassing the resolution limit in optical microscopy[10] or enhancing the nonlinear optical interactions driven in metamaterials or individual nanophotonic systems.[11−13] The spatial and spectral distributions of the induced electromagnetic near-fields determine the functionality of nanophotonic devices. However, characterization of the near-field properties often rely on numerical simulations, which typically represent approximate solutions.

Optical near-fields can be experimentally visualized on their natural length and time scales using photon-induced near-field electron microscopy (PINEM) utilizing the electrons inelastically scattered by the interaction with the localized electromagnetic mode to form an image of the near-field distribution (more precisely the distribution of the Lorentz force integrated along the electron trajectory). Since its invention,[14−17] this technique has been applied for imaging biological structures,[18−20] plasmonic excitations[21−24] or optical near-fields of nanostructures.[25−28] The electron-photon coupling can be understood as a harmonic phase modulation at the light frequency imprinted on the electron wave function.[15] In the particle picture, the electron absorbs or emits individual photons of the coherent excitation field in a stimulated manner. Such interaction is prohibited in vacuum due to different dispersion relations of electrons and photons preventing to fulfill the energy and momentum conservation laws. When a photon is spatially confined at a distance shorter than its wavelength, its momentum becomes delocalized due to Heisenberg uncertainty principle (the same applies for classical fields due to the Fourier-conjugate relationship between spatial localization and wave vector spread) and both the momentum and energy are conserved during inelastic scattering of electrons. Because this process is stimulated, its probability is many orders of magnitude larger than the probability of spontaneous electron energy loss, making PINEM a more sensitive alternative to imaging of the localized photon density of states based on electron energy-loss spectroscopy.[19]

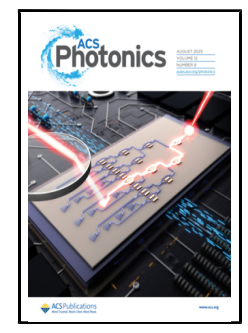

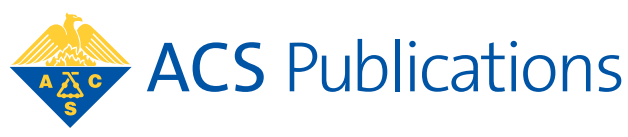

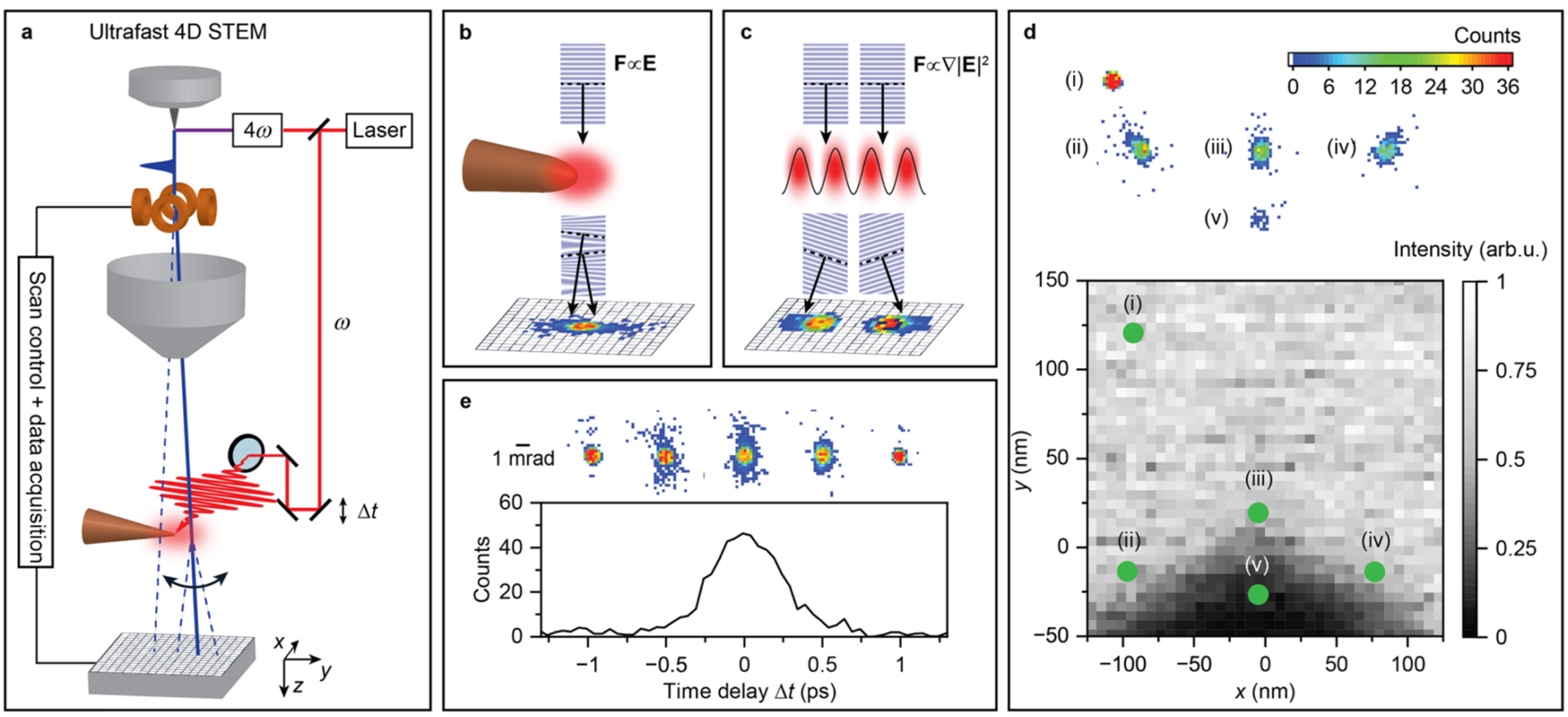

**Figure 1.** Ultrafast 4D scanning transmission electron microscopy (U4DSTEM). (a) Layout of the U4DSTEM experimental setup. (b, c) Illustration of electron phase-fronts in $y-z$ plane in the case of the interaction with (b) a localized optical near-field and (c) the ponderomotive potential of an optical standing wave. (d) Gray scale: scanning transmission electron image of the tungsten nanotip. Color scale shows the measured electron scattering patterns in five positions marked by green points labeled as (i)−(v). (e) Evolution of the number of electrons elastically scattered by the interaction with the optical near-field of a nanotip as a function of the mutual time delay $\Delta t$ between electron and light pulses. Color scale shows the electron scattering patterns measured in time delays of the electron pulse with respect to the laser pulse of $\Delta t = -1{,}25$ ps, $-0{,}3$ ps, 0 ps, 0,4 ps and 1,25 ps in order from left image to right.

Due to a short interaction time of only few femtoseconds in the case of nanostructures with subwavelength dimensions, the electron-photon coupling is rather weak and ultrashort optical pulses with high field amplitudes are required to excite the near-fields. Efficiency of coupling between electrons and localized light modes can be enhanced in the vicinity of periodic nanostructures,[29−32] in evanescent fields leaking from a dielectric crystal[33,34] or from an optical cavity.[35−38] Alternatively to optical near-fields, the electrons can inelastically scatter of optical fields generated at thin membranes[39−42] or via the interaction with optical ponderomotive potential in vacuum.[43−45] The PINEM scheme has been modified by adding a second coherent optical interaction allowing to resolve both the amplitude and phase of the optical near-fields.[46−54] The instrumentation typically used for PINEM-type imaging involves an ultrafast transmission electron microscope with an electron spectrometer or energy filter. In this configuration, the technique is sensitive only to the component of the electric field along the direction of propagation of electrons.[14,15] Because the resulting image does not contain direct information about the local electric field but rather about the total electron momentum change corresponding to the Lorentz force integrated over the interaction distance, it is not possible to reconstruct the full three-dimensional (3D) vectorial electromagnetic field from these images.

Here we propose and demonstrate a novel approach as an alternative to PINEM, enabling the visualization of the interaction between electrons and optical fields in electron microscopes without the need for an electron spectrometer. The method is based on ultrafast 4D scanning transmission electron microscopy (U4DSTEM, experimental setup is shown in Figure 1a, details can be found in Supporting Information, Experimental setup section), by which we monitor the deflection of 20 keV electrons due to the interaction with localized optical fields or with ponderomotive potential of an optical standing wave as a function of beam position in the sample plane. By processing the scattered electron images obtained while scanning the electron beam across the sample we obtain information about the strength and direction of the Lorentz force acting on the electron during the interaction with the optical fields.

The interaction between an electron and electromagnetic fields in vacuum can be described classically, semiclassically, or by a fully quantum approach. The PINEM experiments usually occur in the regime in which the initial electron spectrum is narrower than the photon energy of the excitation. In such a case, the electron coherence time is longer than one temporal period of the oscillating field of the excitation and quantum mechanical interference effects may be expected in the final spectrum. These manifest themselves as discrete peaks separated by the photon energy of the excitation.[14−25] When applying the semiclassical (light is treated classically, electron is described as a wavepacket) and nonrecoil (negligible change of electron momentum during the interaction) approximations, the inelastic electron scattering is a result of time-dependent phase modulation imprinted to the electron wave function. The additional phase acquired by the electron during the interaction with optical fields can be written as[54]

$$\Phi(\mathbf{r}, t) = -\frac{1}{\hbar}\int_{t_0}^{t} \hat{\mathbf{H}}_{\text{int}}[\mathbf{r}(t'), t']\mathrm{d}t' \quad (1)$$

where $\hat{\mathbf{H}}_{\text{int}}[\mathbf{r}(t)] = \{\hat{\mathbf{p}} + e\mathbf{A}[\mathbf{r}(t), t]\}^2/(2m)$ is the interaction Hamiltonian of an electron with charge $e$ and mass $m$ interacting with the electromagnetic field with vector potential $\mathbf{A}(\mathbf{r},t)$ in vacuum and $\mathbf{r}(t)$ is the unperturbed classical trajectory of the electron. The coupling between an electron and the electromagnetic field is described by two terms in $\hat{\mathbf{H}}_{\text{int}}$. The PINEM-type interaction is described by a term linear in the field strength $\approx \mathbf{p_0}.\mathbf{A}(\mathbf{r},t)$, where $\mathbf{p_0} = (0,0,pz)$ is the initial electron momentum and $\mathbf{A}(\mathbf{r},t)$ is the vector potential. The second term responsible for phase modulation of the electron wave is the ponderomotive term $\approx |\mathbf{A}(\mathbf{r},t)|^2$, which becomes important when the interaction occurs in vacuum. When we focus on the former case of linear coupling, the scalar product between the initial electron momentum and the vector potential suggests that only the longitudinal momentum component of the electron is modulated. However, in the optical near-field region, the amplitude of the vector potential changes rapidly in the transverse direction in the $x$−$y$ plane leading to a transverse dependence of the interaction strength. As a consequence, electron phase fronts become tilted with respect to the planar phase fronts of the initial electron wave and the tilt depends on the phase of the oscillating vector potential, which the electron experiences (see Figure 1b). Alternatively, the electron phase fronts may be tilted due to ponderomotive interaction with a spatially modulated optical fields in vacuum (see Figure 1c). The tilted electron waves are deflected in the transverse plane and form a scattered electron image at the detector.

We note that there is one significant difference between the final transverse and longitudinal electron momentum distributions after the interaction with the optical fields. When we consider a strongly localized optical near-field, the electron wave feels periodic oscillations of the vector potential when traveling through. The periodicity of the field in the time domain translates to the periodic phase modulation of the electron wave function in the direction of electron propagation when expressed in the electron's rest frame $z' = z - vt$. It is this periodicity in space and time that leads to the discrete peaks in the longitudinal electron momentum and its kinetic energy spectra. When such periodicity is missing in the transverse spatial direction, no separated peaks are expected in the electron transverse momentum spectra. The coherent peaks are only observed in the case of electrons diffracting on a periodic pattern.[55] We come to the same conclusion when considering the particle picture of the interaction. When photons are localized in the transverse direction at a much smaller spatial scale than the wavelength of the exciting light, they become strongly delocalized in the transverse momentum space. The localization of the field in the $z$-direction in the laboratory frame is removed in the electron's rest frame allowing the appearance of the discrete peaks in the energy and longitudinal momentum spectra.

Since no quantum-mechanical interference effects are to be expected in the transverse electron distribution in the case of localized nonperiodic near-fields under our experimental conditions, we can describe the interaction of the electrons within classical approximation in the electron rest frame, where the $j$ component of the total change of electron transverse momentum at spatial coordinates $x$,$y$ can be expressed as

$$\Delta p_j(x, y) = \int_{-\infty}^{\infty} F_j(x, y, t)\mathrm{d}t \quad (2)$$

where $F_j = e[\mathbf{E} + (\mathbf{v} \times \mathbf{B})]_j$ denotes the $j$-th component of the Lorentz force with the electric and magnetic fields $\mathbf{E} = \mathrm{R}\{\tilde{\mathbf{E}}(\mathbf{r},\omega)g(t - \Delta t)\mathrm{e}^{i\omega t + i\varphi_0}\}$ and $\mathbf{B} = \mathrm{R}\{\tilde{\mathbf{B}}(\mathbf{r},\omega)g(t - \Delta t)\mathrm{e}^{i\omega t + i\varphi_0}\}$, respectively. Here $\Delta t$ represents the electron arrival time and $g(t - \Delta t)$ and $\varphi_0$ are the envelope and phase of the optical fields, respectively. The spatial distributions of the electromagnetic fields $\tilde{\mathbf{E}}(\mathbf{r},\omega)$ and $\tilde{\mathbf{B}}(\mathbf{r},\omega)$ at frequency $\omega$ are obtained from a numerical solution of Maxwell's equations (for details see Supporting Information, section Numerical simulations). The electric and magnetic fields are in the case of quasimonochromatic pulses related to the vector potential via $\mathbf{E} = -\partial\mathbf{A}/\partial t$ (we assume constant electrostatic potential corresponding to zero electrostatic field) and $\mathbf{B} = \nabla \times \mathbf{A}$. When describing the interaction of electrons with the ponderomotive potential in vacuum, the Lorentz force can be integrated over the optical cycle to give the nonrelativistic ponderomotive force $F_j = -\mathrm{e}^2/(4m\omega^2)\nabla j\langle|\mathbf{E}|^2\rangle$.

## RESULTS

**Imaging of Optical Near-Field of a Tungsten Nanotip.** U4DSTEM principle is demonstrated on two model examples. In the first example, we image the transverse component of the optical near-field generated by coherent excitation on the surface of a tungsten nanotip[56] (for details on preparation of the nanotips see Supporting Information, section Preparation of nanotips). The layout of the experimental setup is shown in Figure 1a. A pulsed laser beam with a wavelength of 1030 nm and a pulse duration of 250 fs is incident to the nanotip from the direction perpendicular to the plane defined by the tip symmetry axis and the propagation direction of the electron beam. The excitation light is linearly polarized along the nanotip symmetry axis. The pulsed electron beam with a kinetic energy of 20 keV, repetition rate of 500 kHz and a pulse duration of 500 fs, is focused on the nanotip situated at a working distance of the electron microscope of 11.5 mm. To generate an almost collimated electron beam with a current sufficient for U4DSTEM imaging, we use the highest probe current setting of the electron microscope. The beam divergence is reduced by introducing an objective lens aperture with a diameter of 64 $\mu$m. The resulting electron beam has a spot size in focus of 21 nm and a divergence angle of 1.3 mrad.

The strength of the transverse component of the Lorentz force acting on the electrons is measured as a function of the position of the electron beam in the sample plane by detecting the scattered electron images. In Figure 1d we show the bright-field scanning transmission electron microscopy (STEM) image of the nanotip apex with five selected positions of the electron beam labeled (i)−(v). The images detected by the pixel detector in each of these positions are shown in the upper part of Figure 1d. We observe that the electrons are scattered in the transverse direction by the interaction with the optical near-field. The scattered images carry information about the strength and the direction of the Lorenz force (the main axis of the ellipse rotates according to the preferential polarization of the electric component of the near-field).

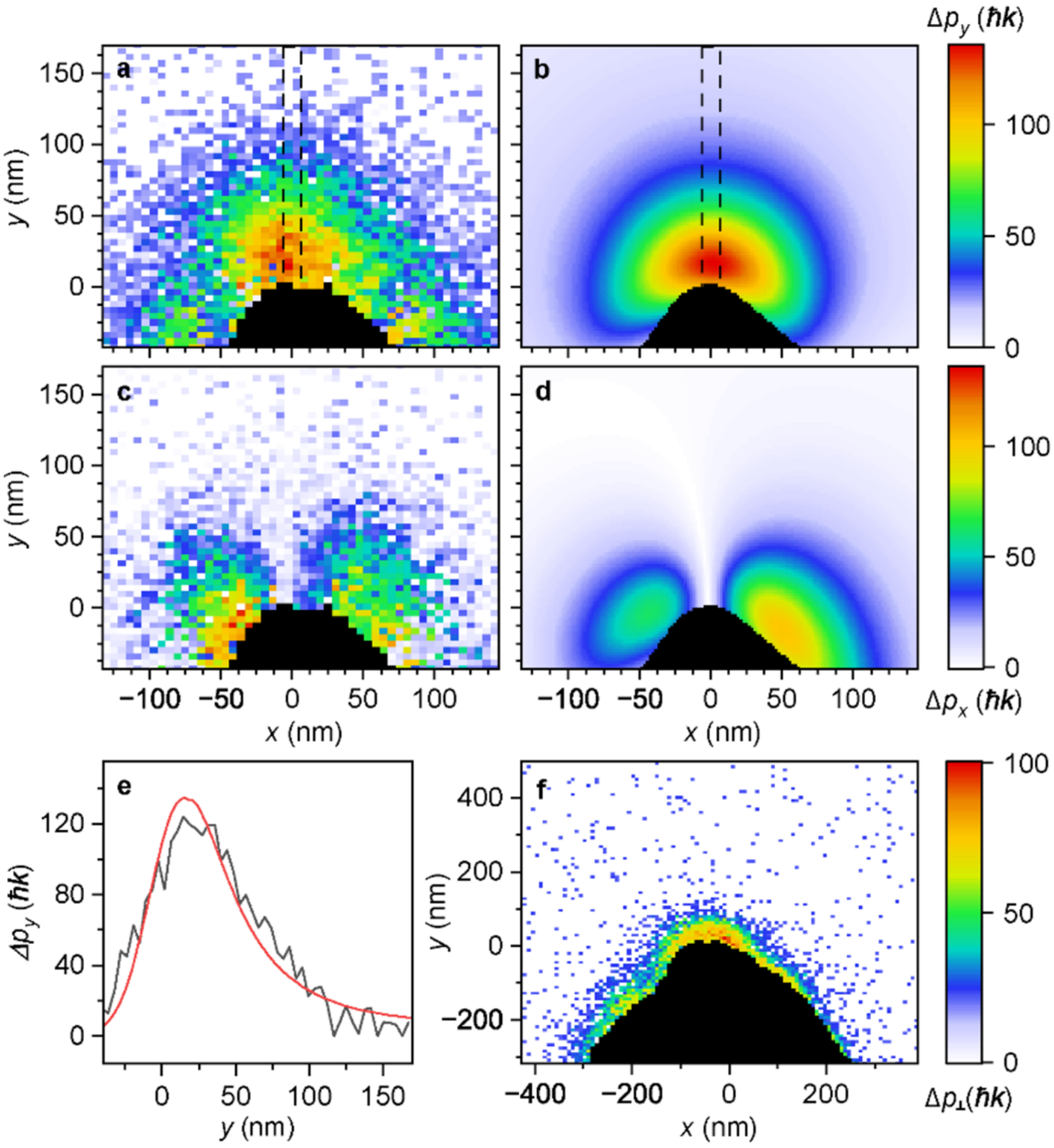

**Figure 2.** Imaging of the transverse component of Lorentz force of optical near-field generated on the surface of a tungsten nanotip by coherent optical excitation. (a) Maximum of the measured $y$-component of the transverse momentum change of the electrons $\Delta p_y$ and (c) its $x$-component $\Delta p_x$ compared to the numerical results shown in (b, d). (e) Comparison of the measured (black curve) and the calculated (red curve) spatial profiles of $\Delta p_y$ obtained by integrating the data in the region marked by dashed lines in (a, b). (f) Maximum of the total transverse momentum change of the electrons $\Delta p_\perp$ induced by the near-field of a blunter tip with larger radius of curvature of the apex. Panels (a−e) use the same scale.

To rule out a significant contribution from elastic electron scattering caused directly by the nanotip surface, we measure the electron scattering patterns as a function of the time delay $\Delta t$ between the electron and laser pulses with the electron beam position fixed in the spot (iii). The measured population of scattered electrons is shown in Figure 1e as a function of $\Delta t$ (black curve) along with the scattered electron images in five selected time delays (color scale). The data clearly show that electron deflection is observed only when the two pulses arrive at the nanotip at the same time, confirming that elastic electron scattering originates from the interaction with the optical near-field instead of the interaction with the nanotip itself.

The transverse components of the near-field are obtained by processing the U4DSTEM data measured at the optimal time overlap between the electron and light pulses. Because the electrons arrive to the sample at different times within the pulse envelope, they experience different amplitude and phase of the oscillating near-field, which is generated by a femtosecond optical pulse. The maximum transverse momentum change of the electrons corresponding to the amplitude of the Lorentz force can be obtained by analyzing the maximum electron deflection in the particular beam position in the sample plane. However, because of the small amount of electrons detected per pixel, the image processing by fitting the streaked electron images exhibits high noise.

For this reason, we apply an alternative data processing method in which we first numerically determine the function describing the relation between the maximum of the Lorentz force and the total number of electrons deflected out of the detector region illuminated by the undeflected electron beam. The maximum change in transverse momentum of the electrons $\Delta p_\perp$ in each position on the sample is then determined purely by counting the deflected electrons. The azimuthal angle $\alpha$ of the maximum of the Lorentz force acting on the electrons in the $x$−$y$ plane is obtained by fitting each image of the scattered electrons by a two-dimensional (2D) Gaussian function and determining the direction of its main axis. The $x$ and $y$ components of $\Delta p_\perp$ are then $\Delta p_x = \Delta p_\perp \cos(\alpha)$ and $\Delta p_y = \Delta p_\perp \sin(\alpha)$ (details of data processing are described in Supporting Information, section Data acquisition and processing). The measured spatial maps of $\Delta p_x$ and $\Delta p_y$ are shown in Figure 2a,b compared to the numerical simulations shown in Figure 2c,d. The amplitude of the measured transverse momentum change of the electrons allows us to estimate the maximum electric field amplitude at the tip surface to be $E_y^{\max} = 3.1 \pm 0.3$ GV/m. The uncertainty was estimated based on the fact that we only exactly know the shape of the $x$−$y$ projection of the tip while the electron deflection is influenced by the extension of the field in $z$ direction see (Supporting Information, section Numerical simulations and Figure S5) for more details about the assumed shape of the nanotip. The amplitude of the electric field of the excitation pulse in the experiments is $E_0 = 0.7 \pm 0.05$ GV/m giving the field enhancement factor of $\xi_{\mathrm{exp}} = 4.4 \pm 0.5$,[57] which agrees well with the value of $\xi_{\mathrm{sim}} = 4.55$ obtained using numerical simulations (for details see Supporting Information, section Numerical simulations). In Figure 2e we compare the measured (black curve) and numerically calculated (red curve) spatial decay of the transverse momentum change of the electrons $\Delta p_y$ obtained by integrating the data in the region of Figure 2a,b labeled by the dashed lines.

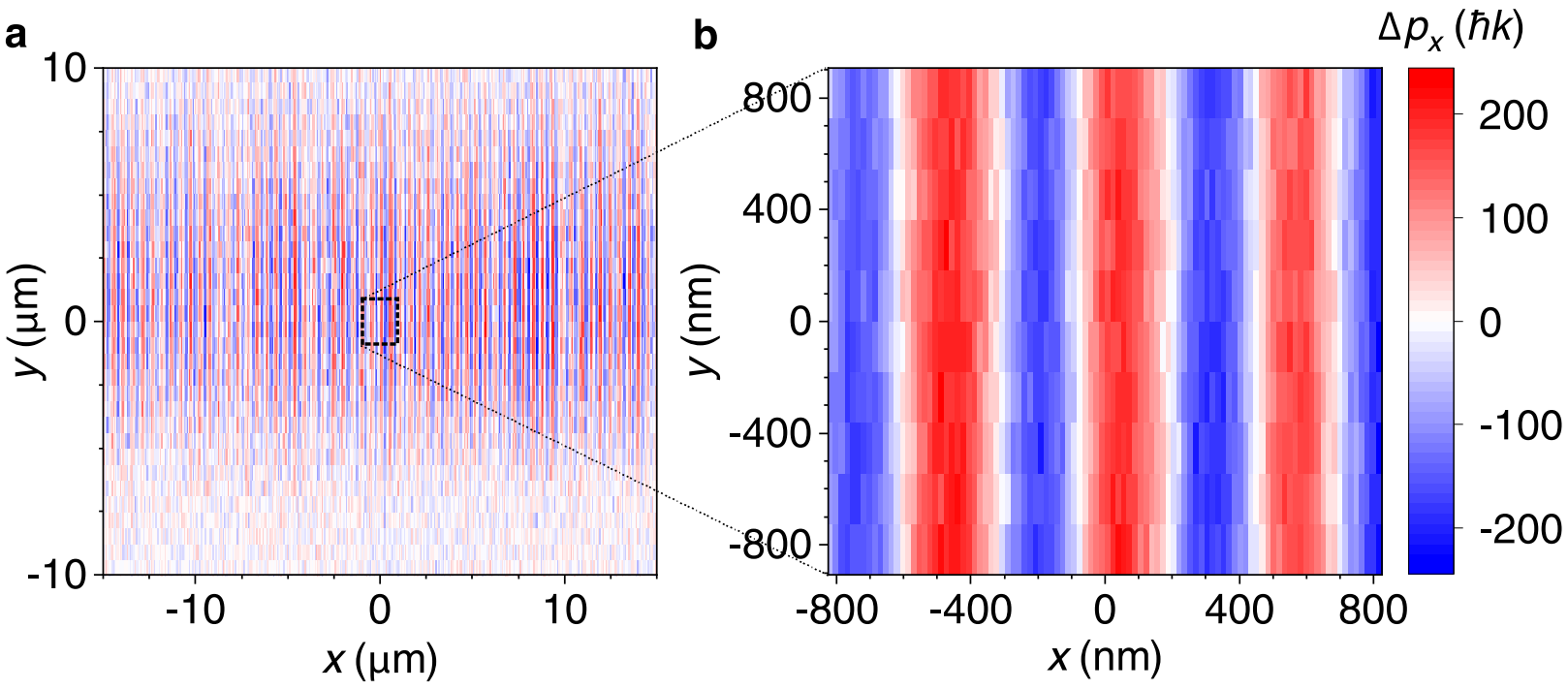

**Figure 3.** Imaging of ponderomotive potential of an optical standing wave in vacuum. (a) Color scale shows the maximum electron momentum change $\Delta p_x$ as a function of the electron beam position in the sample plane. (b) Detail of the image shown in **a**. Both panels use the same color scale.

Bright-field STEM images of the nanotip measured with the pulsed electron beam are used to determine the spatial resolution of the method, which is 21 nm in this experiment. We note that the finite beam width causes the observed shift of the near-field maxima shown in Figure 2e with respect to the tip surface ($x = 0$ nm) because the signal corresponds to the convolution of a sharply decaying function with zero values inside the tip, which describes the near-field amplitude, with the transverse spatial distribution of the electron beam. In Figure 2f we show an image of the amplitude of the integrated transverse optical near-field by measuring the maximum of the total transverse momentum change of the electrons $\Delta p_\perp$ induced by the near-field of a blunter tip with a larger radius of curvature of the apex. From the smaller value of the observed momentum change close to the tip apex and from the fact that the interaction time increases in the case of the blunter tip, we can deduce that the field enhancement is smaller compared to the case shown in Figure 2a. We also examined the role of the polarization state of the excitation light. Data measured with the polarization along the propagation direction of the electron beam (perpendicular to the tip axis) yielded suppressed transverse scattering of the electrons (see Figure S1 in Supporting Information, section Data acquisition and processing).

The amplitude of the electric field estimated from the measured transverse momentum change at the tip apex of about 3 GV/m is expected to lead to nonlinear photoemission of electrons from tungsten surface to vacuum. The quasistatic electric field[58] generated between the positively charged tip and the negatively charged electron cloud may influence the transverse deflection of the electron beam. To study this effect, we use an alternative processing method, in which we characterize the center of mass of the electron distribution on the detector as a function of the position of the beam in the sample plane. We observe that in the time delay $\Delta t = 0$ fs there is a net force deflecting the electrons slightly toward the tip apex in the distance of approximately 50 nm from the tip. When we use a simple model of an electron propagating between two point charges $\pm q$ to estimate the magnitude of the emitted charge from the measured deflection angle (see Supporting Information, section Data acquisition and processing for details), we obtain the value of 40 ± 10 emitted electrons, which is comparable with the number of electrons emitted from a tungsten tip via multiphoton emission observed in experiments performed under similar conditions.[59,60]

**Imaging of an Optical Standing Wave.** In addition, we demonstrate the capabilities of U4DSTEM imaging for visualization of the ponderomotive potential of optical fields in vacuum. For this purpose, we generate an optical standing wave by two counter-propagating pulsed light beams of the same frequency $\omega$. The time-averaged intensity of the electric field can be written as

$$\langle |\mathbf{E}(\mathbf{r},\, t)|^2 \rangle = 2E_0^2[1 + \cos(2k_x x)]G(\mathbf{r},\, t) \tag{3}$$

where $E_0$ is the electric field amplitude of each of the two pulses, $k_x = \omega/c$ is the length of the wave vector corresponding to the optical field forming the standing wave and $G(\mathbf{r},t)$ is the spatiotemporal envelope of the optical beams. The quiver motion of the electrons in the oscillating electromagnetic field of light generates ponderomotive potential. The gradient of the potential expressed by eq 3 along the transverse directions with respect to the electron beam gives an effective force with nonrelativistic expression[61]

$$\begin{aligned} \mathbf{F}(\mathbf{r},\, t) &= -\frac{e^2}{4m_0\omega^2}\nabla\langle |\mathbf{E}(\mathbf{r},\, t)|^2 \rangle \\ &= \left[\frac{e^2E_0^2k_xG(\mathbf{r},\, t)}{m_0\omega^2}\sin(2k_x x),\, 0,\, 0\right] \end{aligned} \tag{4}$$

where we assume a slowly varying envelope approximation $|\nabla G(\mathbf{r},t)| \ll |k_x G(\mathbf{r},t)|$. The U4DSTEM image of the optical standing wave is shown in Figure 3a,b, where we plot the maximum electron momentum change $\Delta p_x$ as a function of the electron beam position in the sample plane. Due to the high gradient of the optical intensity in $x$-direction, the electrons are deflected depending on their position with respect to the standing wave. From the maximum measured momentum change $\Delta p_x$ corresponding to the trajectories of electrons deflected by the largest angles we estimate the peak intensity of the optical standing wave to be 6.52TW·cm$^{-2}$ (see Supporting Information, section Data processing—imaging of the optical standing wave for details). The optical standing wave was imaged by scanning the focused electron beam with its focus located in the plane of the standing wave. Since the dimensions of the electron beam focus are much smaller than the period of the standing wave, the electron beam experiences no spatial periodic phase modulation which would lead to Kapitza-Dirac type scattering[62] or phase shifts measured in previous works.[38] In contrast, we observe position dependent deflection corresponding to the local ponderomotive force.

## ■ DISCUSSION

In principle, the method is also suitable for imaging continuous evanescent fields of resonant structures where absorption/emission of several hundreds of photons by a single electron has been demonstrated in classical PINEM experiments.[37] The spatial resolution of U4DSTEM is in our case limited by the probe size generated by the objective lens in a working distance of approximately 15 mm. To obtain sufficient signal/noise for determination of the integrated Lorentz force we need to detect approximately 50−100 electrons per pixel in the sample plane. To reach this value at the repetition rate of our experiment of 500 kHz we needed to use the objective lens aperture with the size of 64 um, which in combination with the electron lens aberration limits the spatial resolution. The resolution may be improved in the future by implementing higher repetition rates allowing to generate a highly collimated pulsed electron beam less affected by the aberrations of the objective lens, with a smaller spot size while keeping the average electron current at the same level.

In general case the sensitivity of U4DSTEM imaging is limited by the minimum electron beam deflection angle, which can be resolved, unless other case specific data processing methods are suitable (see Supporting Information, section Data processing—imaging of the optical standing wave for details). For our experimental conditions, this limit is posed by the angular size of individual detector pixels, which is 0.33 mrad. The corresponding minimum electric field amplitude that can be detected on the tip surface is $E \approx 1$ GV/m. We note that this number can be significantly smaller for phase-matched structures extended in the electron propagation direction, where the coupling between electrons and photons is strongly enhanced. The detection sensitivity may be further influenced by elastic electron scattering from the sample itself and by the background signal coming from the excitation light scattered to the detector. Although hybrid pixel detectors are not sensitive to individual low-energy photons, when the nanostructure is excited by a femtosecond optical pulse with high peak intensity, many scattered photons can be incident on each pixel within a detection time window of a few nanoseconds. The background optical signal can be avoided by deflecting the electrons by a weak electric or magnetic field and blocking the photons incident on the detector. Alternatively, the optical background can be fully mitigated by using infrared light with photon energy below the band gap of silicon, which forms the active layer of hybrid-pixel detectors.

In conclusion, the U4DSTEM represents an alternative technique to PINEM, which allows us to image the transverse component of the integrated Lorentz force of the optical near-field excited in the vicinity of a metallic nanotip by coherent optical radiation and the ponderomotive potential of an optical standing wave. U4DSTEM can be utilized to image the local density of optical modes of various types of nanophotonic and plasmonic structures, metamaterials or photonic crystals. Combined with the possibility of tuning the frequency of coherent optical excitation, this technique may provide spectral resolution similar to electron energy gain spectroscopy.[63] It offers two significant advantages compared to PINEM. First, it does not require an electron spectrometer, which together with a suitable electron detector represents significant costs in ultrafast electron microscope setup. Second, the U4DSTEM technique can be implemented in low-energy scanning electron microscopes, making it more accessible to users.

## ■ ASSOCIATED CONTENT

### Supporting Information

The Supporting Information is available free of charge at https://pubs.acs.org/doi/10.1021/acsphotonics.5c00864.

Additional details of the experimental setup, preparation of nanotips, data acquisition, data processing, and numerical simulations (PDF)

## ■ AUTHOR INFORMATION

### Corresponding Authors

**Petr Koutenský** − *Department of Chemical Physics and Optics, Faculty of Mathematics and Physics, Charles University, Prague CZ-12116, Czech Republic;* orcid.org/0000-0003-4851-2331; Email: petr.koutensky@matfyz.cuni.cz

**Martin Kozák** − *Department of Chemical Physics and Optics, Faculty of Mathematics and Physics, Charles University, Prague CZ-12116, Czech Republic;* orcid.org/0000-0002-6317-7079; Email: m.kozak@matfyz.cuni.cz

### Authors

**Neli Laštovičková Streshkova** − *Department of Chemical Physics and Optics, Faculty of Mathematics and Physics, Charles University, Prague CZ-12116, Czech Republic;* orcid.org/0009-0004-3931-3597

**Kamila Moriová** − *Department of Chemical Physics and Optics, Faculty of Mathematics and Physics, Charles University, Prague CZ-12116, Czech Republic*

**Marius Constantin Chirita Mihaila** − *Department of Chemical Physics and Optics, Faculty of Mathematics and Physics, Charles University, Prague CZ-12116, Czech Republic*

**Alexandr Knápek** − *Institute of Scientific Instruments of the Czech Academy of Sciences, Brno CZ-61200, Czech Republic;* orcid.org/0000-0003-0752-8214

**Daniel Burda** − *Institute of Scientific Instruments of the Czech Academy of Sciences, Brno CZ-61200, Czech Republic*

Complete contact information is available at: https://pubs.acs.org/10.1021/acsphotonics.5c00864

### Author Contributions

M.K. conceived the study. P.K., M.K., K.M., and M.C.Ch.M. performed the experiments and analyzed the data. A.K. and D.B. prepared the nanotips. P.K. and N.L.S. performed the numerical simulations. P.K. and M.K. wrote the manuscript with the contribution from all coauthors.

### Notes

The work reported in this manuscript was submitted to the arXiv preprint server previous to publication in this journal; the associated paper can be found as follows: Koutenský, P.; Laštovičková Streshkova, N.; Moriová, K.; Chirita Mihaila, M. C.; Burda, D.; Knápek, A.; Kozák, M. Ultrafast 4D scanning transmission electron microscopy for imaging of localized optical fields. 2025, arXiv:2502.07338. arXiv.org e-Print archive. https://arxiv.org/abs/2502.07338 (accessed Feb 11th, 2025).

The authors declare no competing financial interest.

## ■ ACKNOWLEDGMENTS

The authors acknowledge funding from the Czech Science Foundation (project 22-13001K), Charles University (SVV-2023-260720, PRIMUS/19/SCI/05, GAUK 90424) and the European Union (ERC, eWaveShaper, 101039339). Views and opinions expressed are, however, those of the author(s) only and do not necessarily reflect those of the European Union or the European Research Council Executive Agency. Neither the European Union nor the granting authority can be held responsible for them. This work was supported by TERAFIT project No. CZ.02.01.01/00/22_008/0004594 funded by OP JAK, call Excellent Research. All of the data that support the plots and the other findings of this study are publicly available at DOI: 10.5281/zenodo.14824588.[64]

## ■ ABBREVIATIONS

PINEM, photon-induced near-field electron microscopy; U4DSTEM, ultrafast 4D scanning transmission electron microscopy; STEM, scanning transmission electron microscopy; fwhm, full width at half-maximum; FDTD, finite-difference time-domain

## ■ REFERENCES

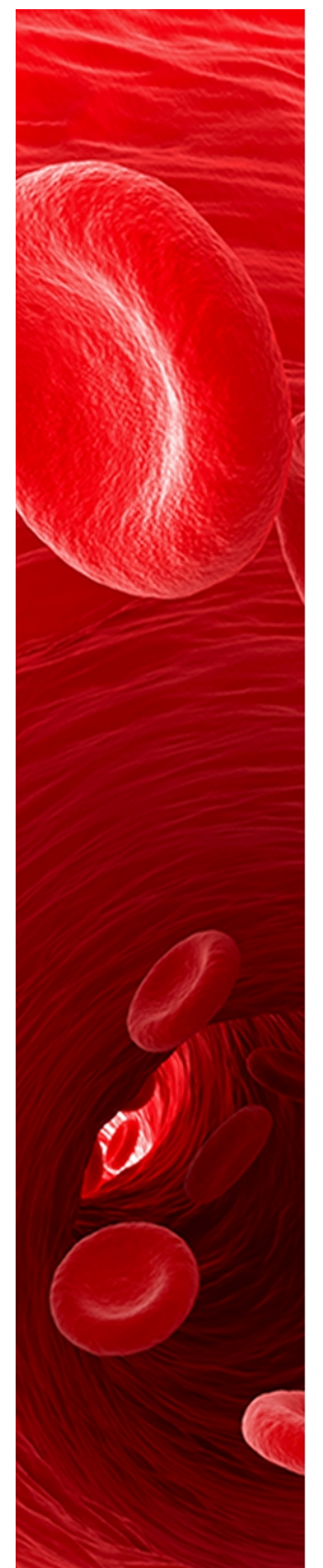

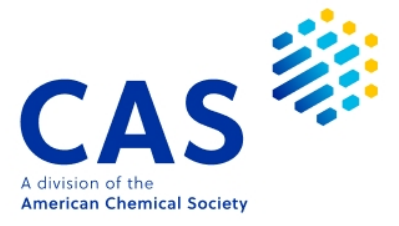